\documentclass[a4paper,12pt]{article}
\usepackage{amsmath}
\usepackage{amssymb}
\usepackage{textcomp}
\usepackage{graphicx}
\usepackage{bm}
\usepackage{color}
\usepackage[dvips]{epsfig}
\def\eps{\epsilon}
\def \l   {\left}
\def \r   {\right}
\newcommand{\p}{\partial}
\newcommand{\ph}{\varphi}
\newcommand{\la}{\langle}
\newcommand{\ra}{\rangle}
\newcommand{\beq}{\begin{equation}}
\newcommand{\eeq}{\end{equation}}
\newcommand{\myref}[1]{~{(\ref{#1})}}
\newcommand{\mycite}[1]{~{\cite{#1}}}
\newcommand{\myfigref}[1]{~{Fig.~(\ref{#1})}}
\newcommand{\be}{\begin{equation}}
\newcommand{\ee}{\end{equation}}
\newcommand{\ba}{\begin{eqnarray}}
\newcommand{\ea}{\end{eqnarray}}

\newcommand{\dmu}{\partial_{\mu}}

\newcommand{\nc}{\newcommand}
\nc{\bi}{\bibitem}

\def\p{\partial}

\def\text{}
\makeatletter
\newcommand\figcaption{\def\@captype{figure}\caption}
\makeatother

\newcommand{\tr}{\mathrm{tr}}

\begin{document}

\begin{titlepage}
\hfill\parbox{40mm} {\begin{flushleft}  ITEP-TH-32/10\\
MPP-2010-167
\end{flushleft}}
\vspace{1mm}
\begin{center}
{\large \bf Low-Energy Theorems from Holography}

\vspace{1mm}

\textrm{Johanna~Erdmenger$^{\dagger}$, Alexander~Gorsky $^*$,
Petr~N. ~Kopnin$^{*\ddagger}$, Alexander~Krikun$^{*\ddagger}$, and
Andrew~V.~Zayakin$^{\flat
*\natural}$}\footnotemark{}\footnotetext{email addresses:
gorsky@itep.ru, jke@mppmu.mpg.de, kopnin@itep.ru,
krikun.a@gmail.com, Andrey.Zayakin@physik.uni-muenchen.de}
\vspace{3mm}

\textit{$^*$ Institute of Theoretical and Experimental Physics,\\
B.~Cheremushkinskaya ul. 25, 117259 Moscow, Russia}\\
\vspace{1mm}

\textit{$^\dagger$     Max-Planck-Institut f\"ur Physik
(Werner-Heisenberg-Institut),\\ F\"ohringer Ring 6, D-80805
M\"unchen, Germany}\\ \vspace{1mm}

\textit{$^\ddagger$ Moscow Institute of Physics and Technology,\\
Institutsky per. 9, 141 700 Dolgoprudny, Russia}\\ \vspace{1mm}

\textit{$^\flat$ Fakult\"at f\"ur Physik der
Ludwig-Maximilians-Universit\"at M\"unchen und\\
 Maier-Leibniz-Laboratory, Am Coulombwall 1, 85748 Garching,
 Germany}

\textit{$^\natural$ Dipartimento di Fisica, Universit\`a di
Perugia, I.N.F.N. Sezione di Perugia,\\ Via Pascoli, I-06123
Perugia, Italy}

\vspace{0.1cm}

{\bf Abstract}
\end{center}
In the context of gauge/gravity duality, we verify two types of
gauge theory low-energy theorems, the dilation Ward identities and
the decoupling of heavy flavor. First, we provide an analytic
proof of non-trivial dilation Ward identities for a theory
holographically dual to a background with gluon condensate (the
self-dual Liu--Tseytlin background). In this way an important
class of low-energy theorems for correlators of different
operators with the trace of the energy-momentum tensor is
established, which so far has been studied in field theory only.
Another low-energy relationship, the so-called decoupling theorem,
is numerically shown to hold universally in three holographic
models involving both the quark and the gluon condensate. We show
this by comparing the ratio of the quark and gluon condensates in
three different examples of gravity backgrounds with non-trivial
dilaton flow.  As a by-product of our study, we also obtain
gauge field condensate contributions to meson
transport coefficients.
\end{titlepage}

\nopagebreak[4]

\tableofcontents \section{Introduction } On the long road towards
a holographic description of QCD, there are some  milestones
corresponding to exact relations which have to be satisfied also
in any holographic model. These are so called low-energy theorems
\mycite{Novikov:1981xj} (see e.g.\mycite{Shifman:1988zk} for
review). In field theory these are statements which impose
restrictions on the various correlators. The purpose of this work
is to compare holography to field theory by considering the
low-energy theorems concerning one- and two-point functions of a
strongly coupled gauge theory on both sides of the correspondence.
We report nice non-trivial agreement in two important cases: the
dilation Ward identities and the decoupling theorem for the heavy
flavor. Recently the validity of a related class of theorems (QCD
sum rules) was shown holographically in\mycite{Gulotta:2010cu} at
finite temperature. Apart from demonstrating the validity of
low-energy theorems, a particular result of our analysis is a
statement on the IR universality of theories dual to three
scale-dependent backgrounds with non-trivial dilaton flow.

First, we aim at realizing the QCD low-energy theorems explicitly,
for instance \beq \int d^4x \langle T(x) O(0)
\rangle=-\mathrm{dim}(O)\langle O\rangle,\eeq where $T=T^\mu_\mu$
is energy-momentum trace on the boundary. This is trivially
satisfied in the conformal case: The right-hand side is expected
to be zero in a conformal field theory where all condensates
vanish. For an explicit expression for the correlators of
energy-momentum components see e.g.\mycite{Liu:1998bu}. Thus for a
nontrivial test we need a background which is different from AdS
in the IR,  dual to a non-conformal field theory, for instance
with a gluon condensate. There are a number of models which
generalize the original AdS/CFT correspondence to the backgrounds
corresponding to non-vacuum states of $\mathcal{N}=4$ SYM or to
non-conformal and non-supersymmetric theories. We use the
self-dual background by Liu and Tseytlin\mycite{Liu:1999fc} with
non-zero expectation value of the gluon operator $\langle\tr
G^2\rangle$ in this part of our work. To perform the test of
dilation Ward identities, we calculate the two-point correlators
$\langle \tr G^2(x) \tr G^2(0)\rangle$, $\langle \tr G^2(x) \tr
G\tilde{G}(0)\rangle$, $\langle T(x) \tr G^2(0)\rangle$, $\langle
T(x) \tr G\tilde{G}(0)\rangle$, $\langle
T_{\mu\nu}(x)T_{\alpha\beta}(0)\rangle$ in this background.

The analysis of correlators is easily performed for non-zero
frequency. In this way we reproduce the results for transport
coefficients, extending the analysis to the case of the
non-conformal backgrounds considered. First of all, we calculate
the $\eta/s$ ratio of shear viscosity over entropy via $\langle
T_{xy}T_{xy} \rangle$, which was performed for the conformal case
in \mycite{Policastro:2002se,Policastro:2001yc,
Son:2002sd,Policastro:2002tn}. Here we find using suitable
holographic renormalization that condensate corrections to $
\frac{\eta}{s}|_{T\to 0}=\frac{1}{4\pi}$ are absent in the
Liu-Tseytlin background, i.e. the nonzero VEV of the gluon field
strength $\langle\tr G^2\rangle$ does not affect the value of
$\eta/s$.

Secondly, we check the relationship between two-point and one
point functions in gauge theory with fundamental fermions, known
as decoupling relation \beq\langle\frac{\alpha_s}{\pi} \tr
G^2\rangle =-12 m\langle\overline{q}q\rangle \, .\eeq  Fundamental
fermions are introduced in our system via probe $D7$ branes, see
e.g.\mycite{Barbon:2004dq}. The $D7$ branes represent the
fundamental degrees of freedom, being convenient locations for the
fundamental strings to end and to be thus endowed with a global $SU(N_f)$
flavor symmetry in the Maldacena limit.
The length of the strings corresponds to the quark
mass, and the subleading term in the asymptotics of the embedding
coordinates to the condensate.  A non-trivial test of the theorem
considered is possible only for an IR-non-trivial metric.  For
that purpose we use three different dilaton flow backgrounds with
gluon condensate: the self-dual
Liu-Tseytlin background mentioned above \mycite{Liu:1999fc}, the
Gubser--Kehagias--Sfetsos background
\mycite{Gubser:1999pk}, \mycite{Kehagias:1999tr}  and the Constable--Myers
background \mycite{Constable:1999ch}. All of these are examples for non-trivial dilaton
flows. A remarkable universality among the three models and
agreement with standard field theory is observed.

Let us now present the two low-energy theorems discussed in this
paper.

\paragraph{\label{dwi}Dilation Ward Identity.} It was argued
in \mycite{Novikov:1981xj} that the following dilation Ward
identity holds within field theory, \beq\lim_{q\to 0}i\int e^{iqx}d^4x \left\langle
T\left\{\mathcal{O}(x),\frac{\beta(\alpha_s)}{4\alpha_s}\tr G^2(0)
\right\} \right\rangle=(-d)\langle\mathcal{O}\rangle
\left[1+\mbox{mass-dependent terms}\right],\eeq where $d$ is the
canonical dimension of the operator $\mathcal{O}$,
$T\{\cdot,\,\cdot\,\}$ stands for the time ordered product and the
one-loop beta-function is normalized as
$\beta(\alpha_s)=-\frac{b\alpha_s^2}{2\pi}$,
$b=\frac{11}{3}N_c-\frac{2}{3}N_f$. Identities for higher
correlators are also available: \beq i^2\int d^4x d^4y
\left\langle
T\left\{\mathcal{O}(x),\frac{\beta(\alpha_s)}{4\alpha_s}\tr
G^2(y),\frac{\beta(\alpha_s)}{4\alpha_s}\tr G^2(0) \right\}
\right\rangle=(-d)^2\langle\mathcal{O}\rangle\left[1+
\mbox{mass-dep.}\right].\eeq

\noindent For the gluon field strength operators we obtain:
\beq\label{scalargluonium} i\int \left\langle
T\left\{\frac{3\alpha_s}{4\pi}\tr G^2(x),\frac{3\alpha_s}{4\pi}\tr
G^2(0)\right\}\right\rangle=\frac{18}{b}\left\langle
\frac{\alpha_s}{\pi}\tr G^2\right\rangle.\eeq

\paragraph{Decoupling Theorem.}
Novikov, Shifman, Vainshtein and Zakharov derived in
\mycite{Novikov:1981xj} the following equation  for light quarks
by considering  the regularity of the beta function
\beq\label{light}\frac{d}{dm_q}\left
\langle\frac{\alpha_s}{\pi}\tr
G^2\right\rangle=-\frac{24}{b}\langle\overline{q}q\rangle.\eeq
\noindent  This low-energy theorem for heavy quarks is recovered
also in an independent manner in\mycite{Gorbar:1999xi}. Besides,
for heavy quarks the following relation due to Shifman, Vainshtein
and Zakharov holds
\beq\label{decoupling}m\langle\overline{q}q\rangle=-\frac{1}{12}
\left\langle\frac{\alpha_s}{\pi}\tr G^2\right\rangle.\eeq  The
derivation of this equation is found in\mycite{Shifman:1978bx}. It
expresses the continuity of the energy-momentum trace at the
flavor number thresholds of the beta-function. The factors 12 and
24 in the equations above are universal, they do not contain $N_c$
or $N_f$. In this paper, we shown that relation (\ref{decoupling})
holds holographically in the three dilaton-flow backgrounds to
great accuracy.

A related calculation, the holographic derivation of the
Veneziano-Witten formula relating the mass of the $\eta'$ meson
and the topological susceptibility of pure Yang-Mills theory, was
performed in \mycite{Barbon:2004dq}. The holographic conformal
anomaly was previously considered under finite temperature in the
5-dimensional model with a dilaton potential adjusted in such way
that both confinement and the correct UV behaviour of the coupling
are reproduced\mycite{Gursoy:2008za}.

This paper is organized as follows. In Section \ref{recipe} we
describe technicalities related to finding correlators. In Section
\ref{ward} we describe the holographic description of the dilation
Ward identities. Section \ref{dec} contains our main result -- the
derivation of the nonperurbative decoupling of the heavy flavor in
the different dilaton flow models. In the last section we discuss
the importance of having established the decoupling and scaling
theorems holographically. Several necessary facts concerning the
models are collected in Appendix A, while Appendix B concerns the
derivation of the transport properties of the models under
consideration.

\section{Recipes of AdS/CFT\label{recipe}}
For later use, let us briefly review the AdS/CFT prescription for
calculating two-point functions, emphasizing in particular the
derivation of the gauge-fixing and the Gibbons-Hawking term. In
the analysis of the boundary term we follow here very closely the
analysis of\mycite{Liu:1998bu}. A reader familiar with these
technicalities can proceed directly
to the next section.

We consider the general rules for
two-point functions and calculate the matrix of correlators

\beq M_{ij}=\langle O_i O_j \rangle|_{(p)}= \frac{\delta^2
S_{full}}{\delta {\bar{\Phi}_i(p) \delta\bar{\Phi}_j(-p)}}.\eeq

\noindent The standard wisdom on finding Green function of the
fields present is to set the action of the type \beq S_{bulk}=\int
d^4x dz \phi^{\prime 2}g^{zz}\sqrt{g}\eeq \

\noindent out onto the boundary as \beq S_{boundary}=\int d^4x
\phi \phi^\prime g^{zz}\sqrt{g}|_{z\to 0}\,\,\,.\eeq

\noindent The correlator in terms of bulk-to-boundary Green
functions $G(x,z)$ of the field $\phi$ is given by \beq \langle
O(x)O(0)\rangle=G(x,z)\partial_z G(0,z)|_{z=0}.\eeq

\noindent In our case two additional difficulties arise. First,
the correct boundary term should be supplemented by the
Gibbons--Hawking term\mycite{Liu:1998bu}, which makes a theory
defined on manifold with boundary globally
diffeomorphism-invariant. Second, the bilinear action of fields'
fluctuations is non-diagonal, this means that we shall be dealing
with a matrix of Green functions rather than with
separately-treatable ones.

\noindent Let us define Green function matrix. Namely, if field
$\Phi_i$ has a bulk solution $\Phi_i(z)$,  satisfying
$z^{\delta_i}\Phi_i(z)|_{z\to 0}=\bar{\Phi}_i$, then by definition
\beq\label{gbf}
K_{ij}(z)=\frac{\delta\Phi_j(z)}{\delta\bar{\Phi}_i}.\eeq

\noindent Let us establish the correct boundary term. The full
action of our bulk theory is actually\mycite{Liu:1998bu} \beq
S_{full}=S_{10d}+S_{div}+S_{4d}\eeq where the Gibbons--Hawking
term\beq S_{4d}= -2
\partial_z\int d^4x \sqrt{-g_4}-c\int d^4x  \sqrt{-g_4}, \eeq
is here given by \beq g_4=\det(g_{ij}),\,\, i=0,1,2,3.\eeq The
constant $c$ can be fixed arbitrarily to our convenience, e.g. as
in eq. (4.15) in\mycite{Liu:1998bu}. The other piece which one has
to take into account is the full divergence term $S_{div}$, which
does not affect equations of motion, but does change the
appearance of the action and makes it diagonal in terms of
physical degrees of freedom of the graviton. It is the well-known
fluctuation term \beq S_{div}=\frac{3}{2}\partial_\mu W^\mu, \eeq
the vector $W^\mu$ is (see \mycite{Landau}, Vol.II, \S 96) \beq
W^\mu = \sqrt{-g}\left(g^{\alpha\beta}\delta
\Gamma^\mu_{\alpha\beta}-g^{\alpha\mu}\delta
\Gamma^\beta_{\alpha\beta} \right),\eeq

\noindent where
$\delta\Gamma^\mu_{\alpha\beta}=\Gamma^\mu_{\alpha\beta}(g+h)
-\Gamma^\mu_{\alpha\beta}(g)$. This constitutes the gauge-fixing
prescription for our problem.

\noindent Consider now the second variation of these actions in
fluctuation fields; denote these second-order expressions as
$S^{(2)}_{10d}$, $S^{(2)}_{div}$, $S^{(2)}_{4d}$ respectively;
they contain both fields and their derivatives. The two-point
correlator is then

\beq\label{bascor} \langle O_i O_j\rangle=K_{ik}\frac{\partial^2
\mathcal{L}}{\partial\Phi_k^\prime
\partial\Phi_m^\prime}\partial_z K_{jm}
+ K_{ik} \frac{\partial^2 {S^{(2)}_{4d}}}{\partial\Phi_k
\partial\Phi_m^\prime} \partial_z K_{jm}
+ K_{ik}\frac{\partial^2 {S^{(2)}_{4d}}}{\partial\Phi_k
\partial\Phi_m} K_{jm},
\eeq here $\mathcal{L}$ is Lagrangian density of the bulk action:
\beq S_{bulk}=S^{(2)}_{10d} + S^{(2)}_{div}=\int dz \,
\mathcal{L}.\eeq

\noindent The above structure is obvious from the following
reasons. Consider the bulk action \beq\delta^2 S_{bulk}
=\frac{\delta\Phi_m(z) }{\delta \bar{\Phi}_j} \frac{\delta^2
S_{bulk}}{\delta \Phi_m\delta \Phi_k } \frac{\delta\Phi_k(z)
}{\delta \bar{\Phi}_i},\eeq where

\beq \frac{\delta^2 S_{bulk}}{\delta \Phi_m\delta \Phi_k }=\int dz
\left[ \frac{\partial^2 L}{\partial \Phi_m^\prime \partial
\Phi_k^\prime }\partial_z\delta\Phi_m\partial_z\delta\Phi_k
+\frac{\partial^2 L}{\partial \Phi_m \partial \Phi_k^\prime
}\delta\Phi_m\partial_z\delta\Phi_k +\frac{\partial^2 L}{\partial
\Phi_m \partial \Phi_k }\delta\Phi_m \delta\Phi_k\right].\eeq

\noindent Taking into account that Green functions of field
fluctuations by definition satisfy equations:

\beq\label{secord} \left[ -\partial_z\frac{\partial^2 L}{\partial
\Phi_m^\prime
\partial \Phi_k^\prime
}\partial_z +\frac{\partial^2 L}{\partial \Phi_m \partial
\Phi_k^\prime }\partial_z +\frac{\partial^2 L}{\partial \Phi_m
\partial \Phi_k }\right]\delta\Phi_k(z)=0,\eeq

\noindent one sees that the only contribution of $S_{bulk}$ into
the correlator will be, after taking off the derivative and
integration, the term:

\beq\delta^2 S_{bulk}= \delta \Phi_m (z)\frac{\partial^2
L}{\partial \Phi_m^\prime
\partial \Phi_k^\prime}\partial_z \delta\Phi_k(z).\eeq

\noindent Now remembering the definition of Green function matrix
\beq K_{mj}=\frac{\delta\Phi_m(z) }{\delta \bar{\Phi}_j},\eeq we
arrive exactly at\myref{bascor}. Then there is the purely boundary
term (Hawking-Gibbons term). It does not require the above
procedure, since it already sits on 4d. Then it contributes the
following: \beq\delta^2 S_{4d}=\frac{\partial^2 S_{4d}}{\partial
\Phi_m^\prime
\partial \Phi_k} \partial_z\delta\Phi_m \delta\Phi_k+
\frac{\partial^2 S_{4d}}{\partial \Phi_m
\partial \Phi_k} \delta\Phi_m \delta\Phi_k.\eeq
The action $S_{4d}$ contains no more than one derivative term,
which is due to normal differentiating of extrinsic curvature,
thus $\frac{\partial^2 L}{\partial \Phi^{\prime 2}}=0$. This
contributes the other two terms into the correlator\myref{bascor}.

\section{Low-Energy Theorems\label{ward}}
In this Section we calculate the matrix of the two-point
correlators for the gluonic operators and components of the
energy-momentum tensor. Then we compare these to one-point
correlators and find that the correct scaling relations from field
theory are satisfied on the gravity side. We begin by introducing
the Liu--Tseytlin model in which we will perform our calculations
in this section.

\paragraph{Liu--Tseytlin model.}\noindent
In the Einstein frame the bulk action of  $IIB$ superstring theory is\mycite{Liu:1999fc} 
\beq \label{D10_action} S_{10} = \frac{1}{g_s^2 (2 \pi)^7
\alpha'^4} \int d^{10} x \sqrt{g_{10}} \left( R - \frac{1}{2}
(\dmu \phi)^2 - \frac{1}{2} e^{2 \phi} (\dmu C)^2 - \frac{1}{2}
|F_5|^2 \right), \eeq \noindent where $R$ is the curvature, $\phi$
is dilaton, $F_5$ is 5-form and $C$ is axion.

The Liu--Tseytlin model is a generalized background for holography
those which possesses self-duality. It describes a field-theory
flow from a strongly-coupled conformal theory in the UV to a
theory with condensate $\tr\, G^2$ in the IR. By virtue of
self-duality it is still supersymmetric. However, it possesses a
scale parameter, which makes it closer to real-world physics. The
self-duality is provided by the presence of a non-trivial axion
field. Despite the presence of the scale, it is conformal in the
UV; in the IR the dilaton singularity is determined by the  gluon
condensate $\tr \, G^2$. Within supergravity this background is
understood as ``smeared'' $D(-1)$ brane with a usual stack of
$D3$-branes. Since $D(-1)$ brane is an instanton in 10D, the
resulting 4d theory can be considered as having an instanton-gas
type of vacuum, which is advantageous for QCD purposes. Moreover,
this background is confining (in the sense of Wilson loop linear
behavior at large temporal separation), and the string tension is
proportional to the condensate. Of course, we do not claim to produce
any real QCD results in this framework, but we believe it to be a
very useful toy model.

For the Liu--Tseytlin background\mycite{Liu:1999fc} metric in
Einstein frame looks like the standard conformal solution
\beq\label{bcgr} ds^2=g_{0\mu\nu}dx^\mu
dx^\nu=R^2\left(\frac{dx^{\mu 2}}{\sqrt{h_3}}\, +\, \sqrt{h_3}\,
\frac{dz^2+z^2 d\Omega_5^2}{z^4}\right), \eeq but the dilaton is
modified by the smeared instanton (nonzero density of $D(-1)$)
\beq e^\phi=h_{-1},\eeq and an axion is present \beq
C_0=\frac{1}{h_{-1}}-1;\eeq the $D3$ and $D(-1)$ form-factors are:
\beq h_3=z^4,\eeq and \beq\label{smear} h_{-1}=1+qz^4.\eeq The
parameter $q$ is the crucial quantity for us, since it measures
the degree of IR-non-conformality of the theory (remember that in
the UV, the theory is conformal and its $\beta$-function is zero).

The Tseytlin-Liu background has been successfully used for a
number of applications, e.g. calculating meson
spectra\mycite{Brevik:2005fs,Erdmenger:2007vj,
Erdmenger:2007cm,Rust:2010gq}. In all these applications, its
relevance to QCD has been demonstrated. In \mycite{Ghoroku:2005tf}
a finite-temperature extension of the\mycite{Liu:1999fc} solution
has been found, which has been a further motivation to apply it to
realistic high-energy quark-gluon plasmas. We shall employ
Liu-Tseytlin background to test dilation Ward identities in
Section \ref{dwi} and decoupling relation in Section \ref{dec}.

\paragraph{Holographic normalization of the
operators\label{normal}}\noindent

Here we consider normalization of the gluon field strength
operator; the normalization of the quark operators will be considered
in the next Section. According to the AdS/CFT dictionary we state
that the fluctuation $\delta \phi(z,Q)$ of dilaton field
\beq\phi(z,Q) = \phi_0(z) + \delta \phi(z,Q)\eeq is dual to the
operator $O_\phi$, proportional to the QCD scalar gluonic operator
\beq \tr ( G^2) \equiv \frac{1}{c_\phi} O_\phi\, .\eeq We can fix the
normalization constant $c_\phi$ by comparing the two-point
functions

\beq \la O_\phi O_\phi \ra = c_\phi^2 \la \tr ( G^2) \tr ( G^2)
\ra .\eeq

\noindent At large momenta the leading behavior of gluonic
correlator in QCD is\mycite{Kataev:1981gr}:

\beq \la \tr (G^2)(Q) \tr (G^2)(Q) \ra = \frac{N_c^2 -1}{4 \pi^2}
Q^4 \ln(Q^2 \epsilon^2) .\eeq

\noindent To obtain a two-point  function from holography we take
the second variation of the action computed on a classical
solution. In the vicinity of the boundary of $AdS_5$ the action
(\ref{D10_action}) for the fluctuation is:

\beq \label{dilaton_action}
 S_{5} = \frac{\pi^3 R^8}{g_s^2 (2 \pi)^7 \alpha'^4} \int d^4
x dz \frac{1}{z^3} \frac{1}{2} \l[-(\p_z \delta \phi)^2 - \p_\mu
\delta \phi \p^\mu \delta \phi + 2 e^{2 \phi_0} \delta \phi (\p_z
C)^2  \r].\eeq

\noindent Here we have taken the near boundary limit $r \gg L$ (so
that $r^2 \simeq \rho^2$) and changed coordinates $z = \frac{R^2}{r^2}$.
$\pi^3$ is the volume of the $S_5$ sphere, $R^8$ came from the
determinant of the metric ($\sqrt{g} = \frac{R^{10}}{z^5}$). The
last term containing the profile of axion field is negligible at
the boundary (small z) because $\p_z C(z) \sim z^3$. We can find
the bulk-to-boundary propagator of $\phi(z,Q)$ at small $z$ and
large $Q^2$. It is

\beq \ph (z,Q)  = \frac{Q^2 z^2}{2} K_2 (Qz), \qquad \ph(0,Q)=1,
\eeq

\noindent where $K_i$ is  McDonald function of the second kind. Now we can compute the second variation of the action.
It is

\noindent \beq \la O_\phi O_\phi \ra = \frac{\delta^2
S_{cl}}{\delta \phi_0 \delta \phi_0} = \left.\frac{\pi^3
R^8}{g_s^2 (2 \pi)^7 \alpha'^4} \frac{1}{2} \ph(z,Q) \frac{\p_z
\ph(z,Q)}{z^3} \right|_{z=\epsilon} = \frac{N_c^2 }{ 4 (2 \pi)^2 }
\frac{1}{8} Q^4 \ln(Q^2 \epsilon^2), \eeq

\noindent where we used the definition $R^4 =4 \pi g_s \alpha'^2
N_c$ and the asymptotic of McDonald function. Comparing this
result with the expression of QCD we find

\beq O_\phi = \frac{1}{4 \sqrt{2}} \tr (G^2) \label{37}.\eeq

\noindent To establish a relation between gluon condensate and the
expansion coefficient of the dilaton field we compute the vacuum
expectation value of $O_\phi$ at zero momentum taking the first
variation of the action with respect to the boundary value of the
field $\phi_0$. At zero momentum near the boundary the dilaton
field behaves as \beq \phi(z) = \phi_0 + \phi_4 z^4.\label{p0}\eeq
For the dual operator given by\myref{37} we find

\beq \la O_\phi \ra = \frac{\delta S_{cl}}{\delta \phi_0} =\left.
\frac{\pi^3 R^8}{g_s^2 (2 \pi)^7 \alpha'^4} \frac{1}{2} \ph(z,Q)
\frac{\p_z \phi(z,Q)}{z^3}\right|_{z=\epsilon} = \frac{N_c^2 }{ 4
(2 \pi)^2 } \,\, 4 \phi_4 .\eeq

\noindent From\myref{37} and\myref{p0} we get the expression for
the gluon condensate \beq\label{c1}\la \tr (G^2) \ra \equiv 4
\sqrt{2} O_\phi = N_c^2 \frac{ 4 \sqrt{2} }{ (2 \pi)^2 } \phi_4
.\eeq

In the Liu-Tseytlin model the infinitesimal fluctuations  of the
fields on the bulk couple to the operators $\tr G^2$, $\tr
G\tilde{G}$, $T_{\mu\nu} $ in the boundary $\mathcal{N}=4$ SYM
theory. Moreover, in the Liu-Tseytlin model the dilaton field
behaves as  $e^{\phi} = 1 + qz^4$, so the parameter of solution
$\phi_4$ in (\ref{c1}) equals $q$ and the scalar and pseudoscalar
gluon condensates are nontrivial and equal to the value given
in\myref{c1}, i.e. \beq\langle\tr G^2\rangle=\langle\tr
G\tilde{G}\rangle=N_c^2 \frac{4\sqrt{2}}{(2\pi)^2}q.\eeq

\paragraph{Correlators at Zero Frequency}\noindent  Fluctuation terms are defined as
\beq\begin{array}{l}\displaystyle \phi=\phi_c+\varphi,\\
\displaystyle C=C_0+\xi,\\ \displaystyle g=g_{0\mu\nu}+h_{\mu\nu}.
\end{array}\eeq

\noindent We consider the following interaction term to provide a
correspondence with the boundary theory:

\beq S_{int}=\int d^4x \left[\frac{1}{2}T_{\mu\nu}\bar{h}^{\mu\nu}
-e^{-\phi_c}  \left(\bar{\varphi} \ \tfrac{\tr G^2}{4 \sqrt{2}} +
\bar{\xi} \ \tfrac{\tr G \tilde{G}} {4 \sqrt{2}}
\right)\right],\eeq which, after introduction of useful self-dual
and anti-self-dual components \beq G^{\pm}=\frac{G\pm
\tilde{G}}{2}\eeq and splitting axion and dilaton fluctuations
into a new couple of variables \beq \eta^\pm =\varphi \pm\xi,\eeq
becomes \beq S_{int}=\int d^4x
\left[\frac{1}{2}T_{\mu\nu}\bar{h}^{\mu\nu} -\frac{e^{-\phi_c}}{4
\sqrt{2}}\left(\bar{\eta}^+\tr G^{+2} + \bar{\eta}^-\tr G^{-2}
\right)\right].\eeq Here bars denote four-dimensional sources,
which are proportional to boundary values of five-dimensional
fields:

\beq \bar{h}_{\mu\nu}=z^2 h_{\mu\nu}|_{z=0},\,
\bar{\eta}^\pm=\eta^\pm|_{z=0},\,
\bar{\varphi}=\varphi|_{z=0}.\eeq

\noindent Fluctuations of $F_5$ are fully determined by
$h^\mu_\mu$, thus there is no independent source for them.

Let us choose the gauge $h_{5\mu}=0$, $k^\mu h_{\mu\nu}=0$,
$u^{\mu}h_{\mu\nu}=0$, where wave-vector $k=(\omega, 0,0,k)$,
constant vector $u$ is $u=(1,0,0,0)$. We work with five fields:
\beq
\bar{\Phi}_i=(\eta^+,\bar{h}_{11}+\bar{h}_{22},\bar{h}_{11}-\bar{h}_{22},
\bar{h}_{12}, \eta^-),\eeq $i=1,\dots 5$, each coupled to the
corresponding $\mathcal{O}_i$ operator\footnote{Some of these
operators, e.g. the ${\mathcal O}_3$ are not of immediate interest; however,
it costs no additional effort to incorporate them into the
calculation, so we work the correlators out for them as well.}

\beq\label{list1} {\mathcal O}_i=\left(\frac{\tr G^{+2} }{4 \sqrt{2}} ,\frac{1}{8}
T^\mu_\mu, \frac{3}{8}T_{11}-\frac{1}{8}T_{22}
-\frac{1}{8}T_{33}-\frac{1}{8}T_{00}, T_{xy},
\frac{ \tr G^{-2} }{4 \sqrt{2}}\right) \, , \eeq

\noindent with $G^+$ and $G^-$ the self-dual and anti-self-dual parts
of $G$, respectively, via\beq\label{via} S_{int} =\int d^4x dz
\sum_{i=1}^5\mathcal{O}_i \Phi_i.\eeq

\noindent The relevant part of the fluctuation action in the bulk
is \beq\label{liutse} S^{(2), \mbox{double deriv.}}_{10d+div}=\int
d^4x dz \left( \frac{1}{z^3}\Phi_1^{\prime}\Phi_5^{\prime}
+\frac{z}{8}\Phi_2^{\prime 2}+\frac{z}{8}\Phi_3^{\prime
2}+\frac{z}{2}\Phi_4^{\prime 2} \right).\eeq One should not be
mislead by its diagonal structure; besides the diagonal terms with
double derivatives, the full bilinear action contains terms which
make it non-diagonal.

The boundary Gibbons-Hawking action term is \beq\label{hawgib}
S^{(2), \mathrm{derivatives}}_{4d}=\int d^4x \frac{1}{8} \left(4
c\, h_{\text{xy}}(z)^2+16 z h_{\text{xy}}'(z)h_{\text{xy}}(z)+\Phi
_2(z) \left(c\, \Phi _2(z)+4 z \Phi_2'(z)\right)\right).\eeq

\noindent The full system of equations upon Green functions
\myref{secord} in the given background\myref{bcgr}--\myref{smear}
is cumbersome and therefore is given in the Appendix B,
eq.(\ref{fsts}). Note that for the $F_5$ form we always have
$\delta F=-2/r^3 \Phi_2$, which solves automatically the equations
of motion for this field and at the same time retains the
constancy of the Ramond-Ramond flow $\int_{S^5}F_5=N_c$.

It is instructive to start with zero-frequency correlators
(setting $\omega=0$ in\myref{fsts} in Appendix B). Subsequently,
we introduce finite frequencies $\omega$. In this case we find
oscillatory solutions (Bessel functions)\myref{phisolomega}
instead of the rational ones\myref{phisol}. The limit $\omega\to
0$ of the finite frequency result coincides with our previously
found result at $\omega=0$ and thus provides an additional check
of the validity for our procedure.

The solutions\myref{phisolomega} contain ten modes labelled by
coefficients $C_i$, $i=1\dots 10$. One would expect that out of
the ten modes five must be IR finite, yet quite unexpectedly six
there are six IR finite modes $(C_1,C_4,C_5,C_6,C_7,C_9)$, and the
remaining four are infinite. An extra constraint is therefore
necessary to make the Green function matrix\myref{gbf} a
well-defined $5 \times 5$ matrix. We require that the resulting
corelator matrix be symmetric, which is equivalent to the
condition $ C_5=C_6/2$, which removes exactly one redundant degree
of freedom.

The Green function matrix is then (recall that $c_\phi = \frac{1}{4 \sqrt{2}} $): \beq\label{gfm}K_{ij}=
\begin{array}{|l|l|l|l|l|l|}  \hline
& c_\phi \ \tr G^{+2}&\frac{1}{8}T^\mu_\mu&\mathcal{O}_3&
\mathcal{O}_4&\ c_\phi  \ \tr G^{-2}\\ \hline
c_\phi  \ \tr G^{+2}& q z^4-q \epsilon ^4+1 & 0 & 0 & 0 & 0 \\
\hline \frac{1}{8}T^\mu_\mu& 0 & \frac{1}{z^2} & 0 & 0 & 0 \\
\hline
 \mathcal{O}_3 & 0 & 0 & \frac{1}{z^2} & 0 & 0 \\  \hline
 \mathcal{O}_4& 0 & 0 & 0 & \frac{1}{z^2} & 0 \\  \hline
c_\phi  \ \tr G^{-2}& -2 q \left(\epsilon ^4-z^4\right)
 & 0 & 0 & 0 & q z^4-q \epsilon ^4+1 \\  \hline
\end{array}
\eeq with the ${\mathcal O}_i$ as given by \eqref{list1}. $q$ is
the non-conformality parameter defined in \eqref{smear}.

As a result, combining our knowledge of Green function
matrix\myref{gfm}, the boundary action\myref{hawgib} and the
derivative piece of the bulk action\myref{liutse} we obtain the
matrix: \beq M=
\begin{array}{|l|l|l|l|l|l|}  \hline
&c_\phi  \ \tr G^{+2}&\frac{1}{8}T^\mu_\mu&\mathcal{O}_3&
\mathcal{O}_4& c_\phi  \ \tr G^{-2}\\ \hline c_\phi  \ \tr
G^{+2}&-4 q & -2q & 0 & 0 & -2 q \\ \hline
\frac{1}{8}T^\mu_\mu&-2q & -\frac{1}{4 \epsilon ^4} & 0 & 0 & 0 \\
\hline \mathcal{O}_3& 0 & 0 & -\frac{1}{4 \epsilon ^4} & 0 & 0 \\
\hline \mathcal{O}_4& 0 & 0 & 0 & -\frac{1}{\epsilon ^4} & 0 \\
\hline c_\phi  \ \tr G^{-2}& -2 q & 0 & 0 & 0 & 0\\ \hline
\end{array},\eeq
which contains information on the correlators of $\mathcal{O}_i$,
$\mathcal{O}_j$ via the following relation \beq
\langle\mathcal{O}_i\mathcal{O}_j\rangle=\frac{N_c^2}{16 \pi^2}
M_{ij}.\eeq

Some comments are due here. The singular terms $\frac{1}{\eps^4}$
are expected due to the divergencies on the field theory side;
they are subtracted by a holographic renormalization procedure,
analogously to field-theoretical subtraction. The asymmetry in
${\mathcal O}_1\leftrightarrow {\mathcal O}_5$ is also expected:
what we consider is a self-dual configuration, therefore, the
self-dual and the anti-self-dual operators have different
properties.

Using the matrix elements obtained above, we can now establish the
low-energy theorems. After  normalization according to\myref{37}
we have \beq\left\{\begin{array}{rcl}\label{correl}\displaystyle
\int d^4 x\left\langle \tr G^{+2}(x) T(0)\right\rangle
&=&\displaystyle 4\left\langle \tr
G^{+2}(0)\right\rangle,\\ \\
\displaystyle \int d^4 x \left\langle \tr G^{-2}(x)
 T(0)\right\rangle &=&0,\\ \\
\displaystyle \int d^4 x \left\langle \tr G^{2} (x)
\tr G^{2}(0) \right\rangle&= &\displaystyle
\frac{1}{2}\left\langle \tr
G^2 \right\rangle,\\ \\
\displaystyle \int d^4 x\left\langle \tr G\tilde{G}(x)
 \tr G\tilde{G}(0) \right\rangle &=& 0.
\end{array}\right. \eeq
where $ T=T_{\nu\nu}$. Here we see that the first and the second
lines of the equations above\myref{correl} constitute exactly the
statement of the low-energy theorems \beq \langle\hat{O} T
\rangle=\mathrm{dim} (O) \langle\hat{O} \rangle.\eeq \noindent
Note that $\langle \tr \, (G^{-})^2 \rangle=0$.

The third line of\myref{correl} must be compared to the
field-theoretical result \beq\int \langle \tr G^2 \tr G^2\rangle
\sim 1/\beta_0 \langle  \tr G^2\rangle,\eeq where in standard
perturbation theory, $\beta_0$ is the one-loop coefficient of the
beta-function. This equation reflects a breaking of the conformal
symmetry. For the Liu--Tseytlin model the standard beta function
vanishes. Nevertheless, the massive parameter $q$ generates
additional terms in the effective action. This gives rise to the
contribution $T_{\mu\mu}\sim \tr G^{-2}$ to the trace of the
energy-momentum tensor at the operator level. This is consistent
with the low-energy theorem given by the third line
of\myref{correl}. On the other hand, $\langle \tr
G^{-2}\rangle=0$, thus the expectation value of the
energy-momentum tensor and the vacuum energy vanish, ensuring
consistency with supersymmetry.


The fourth relation in\myref{correl} implies that the topological
susceptibility of the vacuum, which is proportional to this
correlator\mycite{Witten:1979vv}, vanishes in the Liu--Tseytlin
model, which is in the agreement with the fact that the model is
supersymmetric\footnote{Note that in the D4/D6
model\mycite{Barbon:2004dq} the topological susceptibility does
not vanish. However there is no contradiction between these facts,
since the model of\mycite{Barbon:2004dq} breaks supersymmetry
(similarly to Sakai-Sugimoto model), whereas Liu-Tseytlin model
retains supersymmetry.}.

\paragraph{Correlators at Finite Frequency} \noindent Now let
us analyze the finite-frequency solutions. The solutions are given
in Appendix, eq.\myref{phisolomega}; only relevant modes shown.
Unlike the $\omega=0$ solutions, which were exact solutions, here
$\Phi_2(z)$ and $\Phi_5(z)$ are powerlog expansions in $\omega$
and $r$. Since we are interested in the near-UV behaviour of Green
functions, and eventually expand correlator matrix in powers of
$\omega$, this approximation is reasonable. The matrix of
correlators becomes: \beq M=
\begin{array}{|l|l|l|l|l|l|}  \hline
& c_\phi \ \tr G^{+2}&\frac{1}{8}T^\mu_\mu&\mathcal{O}_3&
\mathcal{O}_4& c_\phi \ \tr G^{-2}\\ \hline  c_\phi \ \tr G^{+2}&
-4 q & -2 q & 0 & 0 & \frac{
 \log \left(\omega e\right) \omega^4}{8} -2q\\ \hline
\frac{1}{8}T^\mu_\mu& -2 q & -\frac{\log
 \left(\omega e\right) \omega^4}{32} & 0 & 0 & 0 \\ \hline
\mathcal{O}_3& 0 & 0 & -\frac{\log \left(\omega e\right)
 \omega^4}{32} & 0 & 0 \\ \hline
\mathcal{O}_4& 0 & 0 & 0 & -\frac{\log \left(\omega e\right)
\omega^4}{32} & 0
 \\ \hline c_\phi \ \tr G^{-2}&
 \frac{\log (\omega e) \omega^4
 }{8} -2q & 0 & 0 & 0 & 0 \\ \hline
\end{array}
\eeq

\noindent  The most interesting physical implication of this
correlator matrix comes from the $\langle T_{xy} T_{xy}\rangle$
element. It is proportional to $\frac{\eta}{s}|_{T=0}$, and here
we observe its independence of $q$. This fact is not trivial from
dimensional considerations, since we possess another dimensionful
parameter, the frequency $\omega$. Thus we have established \beq
\left.\frac{\eta}{s}\left(q,\omega\right)\right|_{T=0}
=\frac{1}{4\pi}.\eeq As a bonus of this calculation, in the
Appendix~\ref{quarkonia} we easily elaborate the matrix of
quarkonium transport coefficient based on the above correlator
matrix.

\section{\label{dec}Holographic Decoupling of the Heavy Flavor}
\subsection{Physics of Decoupling}
\begin{center}
\begin{figure}[h!]
\begin{center}
\includegraphics[height = 3.1cm, width=4cm]{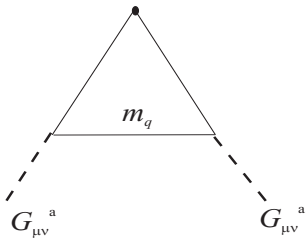}
\caption{\label{triangle} Vacuum diagram with heavy quarks
depicting $\langle\overline{q}q\rangle$ as gluon-driven quantity.}
\end{center}
\end{figure}\end{center}

In this Section we holographically derive the central result of
this paper, which is known as ``decoupling relation''. In can be
found in\mycite{Shifman:1978bx}: \beq
\frac{\alpha_s}{\pi}\left\langle
G^a_{\mu\nu}G^a_{\mu\nu}\right\rangle=-12 m_q
\langle\overline{q}q\rangle.\label{61}\eeq The derivation of this
relation is somewhat intuitive, but let us still restate the
arguments by Shifman, Vainshtein and Zakharov. For vacuum
expectation values of the different operators pertinent to light
quarks the parameter of expansion is quark mass. For heavy quarks
we expand in the inverse quark mass and set external momentum to
$Q^2\sim 0$. Let us suppose there exists a quark for which both
expansions, small and large $m$ are true. As it is in particular a
``heavy'' quark, the quark condensate can be done perturbatively
from the triangle diagram with gluons as ``vacuum sources'', shown
in\myfigref{triangle}.
One can understand the argument from which the
relation\myref{decoupling} emerges as follows. Consider the trace
of energy-momentum tensor of a gauge theory. For low quark mass
there is beta-function contribution from the quark, for heavy
quark there is only the gluonic contribution to the beta-function,
yet there is quark chiral condensate is present:
\beq\theta^\mu_\mu=\left\{\begin{array}{l}\displaystyle
\left(\frac{11}{3}N_c- \frac{2}{3}\right)\frac{\alpha_s}{8\pi}\tr
G^2,
\quad\mbox{above threshold},\\
\\ \displaystyle \left(\frac{11}{3}N_c
\right)\frac{\alpha_s}{8\pi}\tr
G^2+m\overline{q}q,\quad\mbox{below
threshold}.\end{array}\right.\eeq When the two are equated at some
intermediate scale, the necessary relation\myref{decoupling}
appears. Equating small and large $m$ domains happens on the
ground that we select the scale at which the heavy quarks
``decouple'' from the one-loop polarization operator. Hence this
theorem is also known as decoupling relation. A picture of
condensate as function of quark mass is given
in\mycite{Novikov:1981xj}.

\subsection{Decoupling in Specific Backgrounds}  We now
establish relation\myref{61} holographically by considering
different backgrounds, those of Constable and
Myers\mycite{Constable:1999ch}, of Gubser\mycite{Gubser:1999pk}
and of Liu and Tseytlin. The Liu and Tseytlin
background\myref{bcgr} was already discussed above in the Section
\ref{dwi}. The Constable---Myers background in the Einstein frame
has the metric \beq
ds^2=\frac{\left(\frac{b^4+r^4}{r^4-b^4}\right)^{\frac{1}{8
b^4}}}{\sqrt{h_3}} dx_\mu^2+\frac{\left(r^4-b^4\right)
\left(\frac{b^4+r^4}{r^4-b^4}\right)^{\frac{1}{4}
\left(2-\frac{1}{2 b^4}\right)}
\sqrt{\left(\frac{b^4+r^4}{r^4-b^4}\right)^{\frac{1}{2
b^4}}-1}}{r^4}\left( dr^2 + r^2 d\Omega^2_5\right), \eeq where
\beq h_3=\left(\frac{b^4+r^4}{r^4-b^4}\right)^{\frac{1}{2
b^4}}-1,\eeq and the dilaton is \beq
e^\phi=\left(\frac{b^4+r^4}{r^4-b^4}\right)^{\frac{1}{2}
\sqrt{10-\frac{1}{4 b^8}}}, \eeq axion is zero, and
$F_5=\epsilon_5 \frac{1}{h_3}$, where $\epsilon_5$ is the unitary
antisymmetric tensor in the $S_5$ directions.

The chiral condensate and meson spectrum involving a Goldstone
boson were obtained in\mycite{Babington:2003vm} by embedding a D7
brane probe into a Constable--Myers background. Masses of
heavy-light mesons in this background in D7 model were obtained
in\mycite{Erdmenger:2006bg}. The quark condensate, pion decay
constant and the higher order Gasser- Leutwyler coefficients were
calculated for D7 model in this background
in\mycite{Evans:2004ia}. D7 embeddings were argued to be stable in this
background\mycite{Apreda:2005hj,Apreda:2006ie}.

One of the first non-conformal backgrounds introduced into AdS/CFT
was considered by Gubser\mycite{Gubser:1999pk}: \beq
ds^2=\sqrt[4]{1-\frac{b^8}{r^8}} r^2 dx_\mu^2+\frac{1}{r^2}\left(
dr^2 + r^2 d\Omega^2_5\right), \eeq dilaton in this background is
\beq e^\phi=\left(\frac{\frac{r^4}{b^4}+1}{\frac{r^4}{b^4}-1}
\right)^{\sqrt{\frac{3}{2}}}, \eeq

\noindent and the axion is zero. Originally it was intended to
model confinement, yet it became also useful for introducing the
gluon condensate. Shortly before Gubser, this background was also
obtained by Kehagias and Sfetsos\mycite{Kehagias:1999tr} in a less
convenient parametrization.

\begin{figure}[th]
\unitlength=1mm \epsfig{file=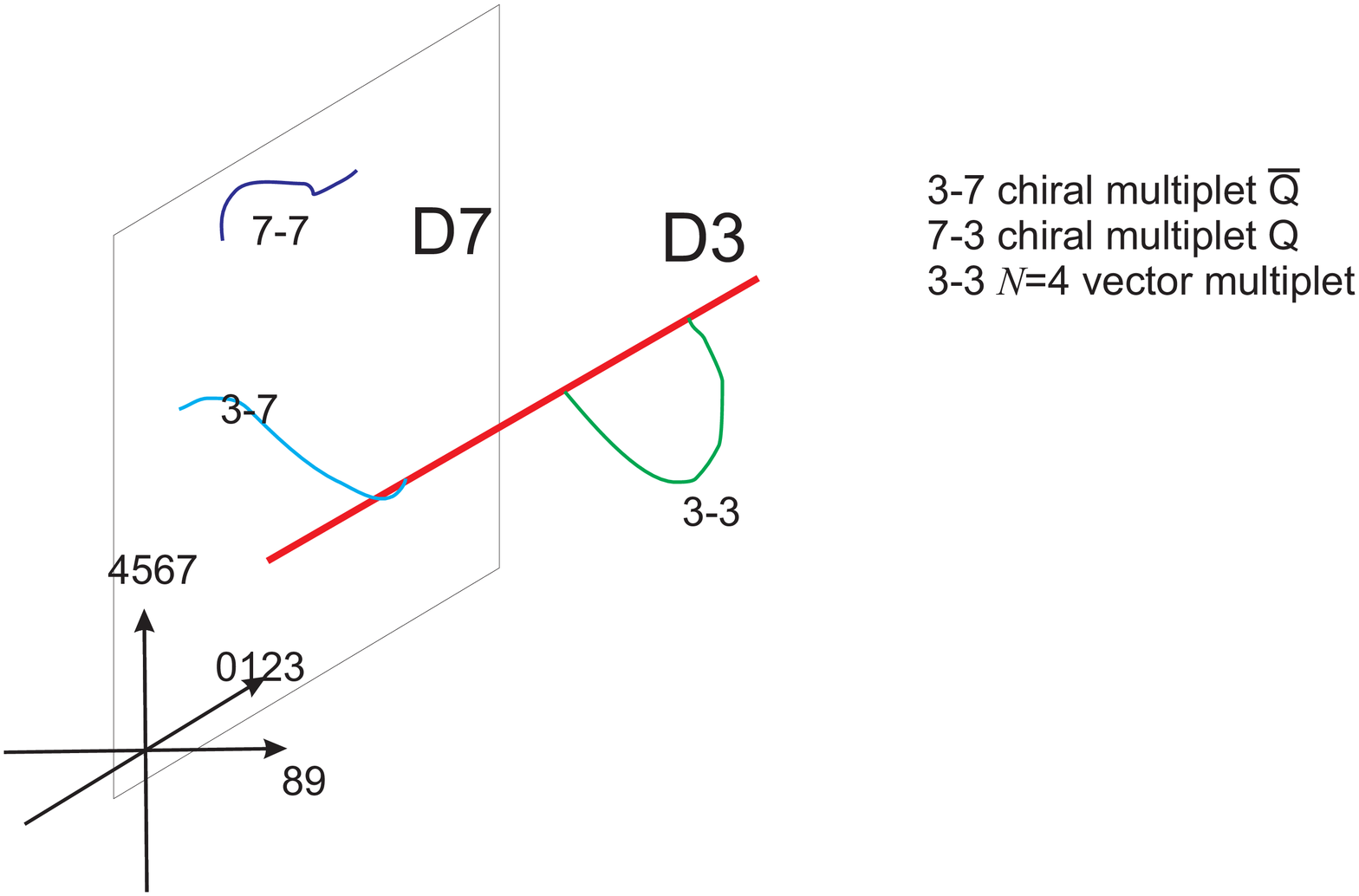,
width=9.0cm,height=5.2cm,angle=0}\label{D3D7}\caption{Scheme of
the D3-D7 geometry and the corresponding string/field modes.}
\end{figure}

\paragraph{Introduction of fundamental fields.}
\noindent We are modelling the fundamental fermionic degrees of
freedom by embedding the D7 brane into one of the three
backgrounds described above. The Dirac--Born--Infeld action for
the $D7$ brane embedding in Einstein frame is given by\beq
\label{D7_action} S_{D7}=\frac{1}{g_s(2\pi)^7\alpha^{\prime
4}}\int d^8\xi \,\,e^{\phi}
\sqrt{\det_{\alpha\beta}\left(\partial_\alpha X^\mu\partial_\beta
X^\nu g_{\mu\nu} \right )}.\eeq The embedding of $D7$ is made as shown
in the following table:
\beq\begin{array}{|l|l|l|l|l|l|l|l|l|l|l|}
\hline AdS_5\times S^5& 0&1&2&3&4&5&6&7&8&9\\
\hline D7&+&+&+&+&+&+&+&+&-&-\\ \hline
\end{array}.\eeq
One can get an image of the corresponding physics
in\myfigref{D3D7}, where string modes generating specific sectors
of the spectrum are shown.

\begin{figure}[h!]
\begin{center}
\includegraphics[height = 4.1cm, width=13cm]{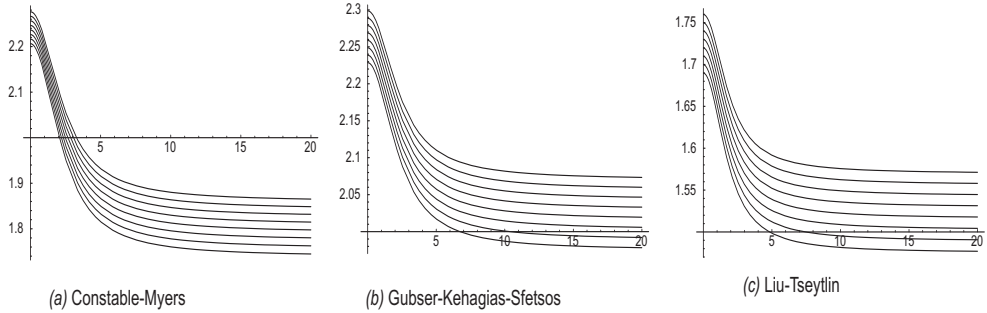}
\caption{\label{embed}Typical embeddings of $D7$ branes.}
\end{center}
\end{figure}

\noindent We look for embeddings of the form \beq X^9=w(\rho),\,\,
X^8=0,\eeq where embedding function $w$, worldsheet coordinates
$\xi_i$ and target space coordinates $r,\rho$ are related as
follows
\beq\begin{array}{l} w^2(\rho)=r^2-\rho^2,\\
\rho=\sqrt{\xi_5^2+\xi_6^2+\xi_7^2+\xi_8^2}.\end{array}\eeq $N_f$
quark flavours can be considered introducing $N_f$ corresponding
$D7$ branes with embedding coordinates $w_i, i=1\dots N_f$. If the
quark masses are equal, $D7$ branes form a stack and the action
(\ref{D7_action}) is multiplied by the factor $N_f$. In the
following we restrict ourselves to the case of just one flavour
for simplicity, considering only one embedding coordinate
$w(\rho)$. Using these definitions we easily construct the
equations of motion for $w(\rho)$, \beq\begin{array}{l} 2 \rho
g_{00}(r) w'(\rho ) \left(w'(\rho )^2+1\right) g_{55}'(r)-2 w(\rho
) \left(w'(\rho )^2+1\right)
\left(g_{55}(r) g_{00}'(r)+g_{00}(r) g_{55}'(r)\right)+\\
+g_{55}(r) \left(2 \rho g_{00}'(r) w'(\rho )^3+2 \rho  g_{00}'(r)
w'(\rho )+r g_{00}(r) w''(\rho )\right)=0, \end{array}\eeq where
the corresponding  $g_{ii}$ should be taken for each respective
metric. We solve them numerically at different values of the
vacuum parameters and fields, corresponding to the boundary
conditions at $\rho\to \infty$; a typical embedding is shown
in\myfigref{embed}.

\subsection{Normalization of the ``Quark'' Operators} \noindent
Following the same steps as in Section \ref{normal} we explore the
scalar field $w$ dual to the   operator $ \bar{q} q$, where $q$ is
the quark field. It is described by the action of the D7 brane
(\ref{D7_action}), for which $w_i$ is embedding coordinate. Here
and after we are dealing only with flavour $i$ and will omit this
index where it is possible. The action for the fluctuations of $w$
is \beq S_{5} = - \frac{2 \pi^2 R^4}{g_s (2 \pi)^7 \alpha'^4} \int
d^4 x dz e^{\phi} \l[ \frac{1}{2 z} (\p_z w)^2 + \frac{1}{2 z}
\p_\mu w \p^\mu w \r]. \eeq

\noindent Here we change coordinates the same way as in
(\ref{dilaton_action}), $2 \pi^2$ is a volume of 3-sphere $R^4$
comes again from the determinant of the metric $\sqrt{g^{(8)}} =
\frac{R^6}{z^3}$. In the limit of large momenta near the boundary
the bulk-to-boundary propagator is

\noindent \beq \tilde{w} (z,Q)  = Q z \ K_1 (Qz), \qquad
\tilde{w}(0,Q)=1. \eeq

\noindent The scalar field is dual to the operator $O_w$, which is
proportional to $ \bar{q} q = \frac{1}{c_w} O_w$. We compute
two-point function of $O_w$ to fix the normalization

\beq\begin{array}{l} \displaystyle \la O_w O_w \ra =
\frac{\delta^2 S_{8 cl}}{\delta w_0 \delta w_0} = \frac{2 \pi^2
R^4}{g_s (2 \pi)^7 \alpha'^4} \ e^{\phi} \frac{1}{2}
\tilde{w}(z,Q) \frac{\p_z \tilde{w}(z,Q)}{z} |_{z=\epsilon} \\ \\
\displaystyle \hphantom{\la O_w O_w \ra } =\frac{N_c}{2 (2 \pi)^4
\alpha'^2} \frac{1}{2} Q^2 \ln (Q^2 \epsilon^2)|_{z=\epsilon}
.\end{array}\eeq

\noindent Here the fact is used that  $e^\phi |_{boundary} = 1$
\mycite{Liu:1999fc} and again $R^4 = 4 \pi g_s \alpha'^2 N_c$. We
compare this result with the QCD calculation (see eq. 4.27
in\mycite{Shifman:1978bx}),

\beq \la \bar{q} q \,\, \bar{q} q\ra =\frac{N_c}{16 \pi^2} Q^2
ln(Q^2 \epsilon^2), \eeq  and find \beq O_w = \frac{1}{2 \pi
\alpha'} \bar{q} q. \eeq

\noindent At this stage we can identify  the boundary value of the
field $w_0 = w|_{z=0}$. It is the source of $O_w = c_w (\bar{q}
q)$, so it is proportional to the quark mass $w_0 =\frac{1}{c_w}
M$. Thus we have \beq \label{m1}M = \frac{1}{2 \pi \alpha'}
w_0.\eeq

\noindent To identify quark condensate  $\la \bar{q}q \ra$ we
compute the expectation value of $O_w$ at $Q=0$. In this limit
near the boundary the supergravity field takes the asymptotic form
\beq \label{w0}w(z) = w_0 + w_2 z^2.\eeq The result is

\beq \la O_w \ra = \frac{\delta S_{8 cl}}{\delta w_0} =
\left.\frac{2 \pi^2 R^4}{g_s (2 \pi)^7 \alpha'^4} \ e^{\phi} \
\frac{1}{2} \tilde{w}(z,Q) \frac{\p_z w(z,Q)}{z}
\right|_{z=\epsilon} =\frac{N_c}{2 (2 \pi)^4 \alpha'^2} 2 w_2.
\eeq

\noindent The quark condensate is normalized as follows \beq
\label{c2}\la \bar{q} q \ra = \frac{1}{c_w} \la O_w \ra =
\frac{N_c}{(2 \pi )^3 \alpha' } w_2. \eeq

\noindent To check the decoupling theorem,  we have to study the
relation $\frac{M \la \bar{q}q \ra}{\frac{g_{YM}^2}{4\pi^2} \la
\tr(G^2) \ra}$ for one specific quark  flavour, i.e. $N_f=1$.
Expressed via the parameters of our model
(\ref{c1}),(\ref{m1}),(\ref{c2}), it turns out to be

\beq \frac{M \la \bar{q}q \ra}{\frac{g_{YM}^2}{4\pi^2} \la
\tr(G^2) \ra} = \frac{ 1}{N_c  4 \sqrt{2}   \alpha'^2
g_{YM}^2} \frac{ w_0 w_2}{\phi_4}, \eeq

\noindent with coefficients $w_0,w_2,\phi_4$ defined
in\myref{p0},\myref{w0}. It is convenient to express all
coefficients via the expansion parameters in the coordinate $r =
\frac{R^2}{z}$. We denote them by $\phi = \phi_0 +
\frac{b_4}{r^4}, w = a + \frac{c}{r^2}$. Obviously, these are
related to the former defined in\myref{p0},\myref{w0} by $\phi_4 =
\frac{b_4}{R^8}, \omega_2 = \frac{c}{R^4}$. Recalling that $R^4 =
4 \pi g_s \alpha'^2 N_c$ and $g_{YM}^2 = 4 \pi g_s$, we obtain

\beq \frac{M \la \bar{q}q \ra}{\frac{g_{YM}^2}{4\pi^2} \la
\tr(G^2) \ra} =
\frac{ 1}{   4 \sqrt{2}   }  \frac{ a c}{b_4}. \label{acb} \eeq

\noindent For the theorem \beq \frac{M \la \bar{q}q
\ra}{\frac{g_{YM}^2}{4\pi^2} \la \tr(G^2) \ra}  =
-\frac{1}{12}\eeq

\noindent to hold, the parameters $a,b_4,c$ must satisfy \beq
\label{sqr}\frac{ac}{b_4} = -\frac{\sqrt{2}}{3}.\eeq This
relation, equivalent to the decoupling theorem, will be tested
numerically below.

\subsection{Numerics}
\begin{figure}[h!]
\begin{center}
\includegraphics[height = 6.1cm, width=9cm]{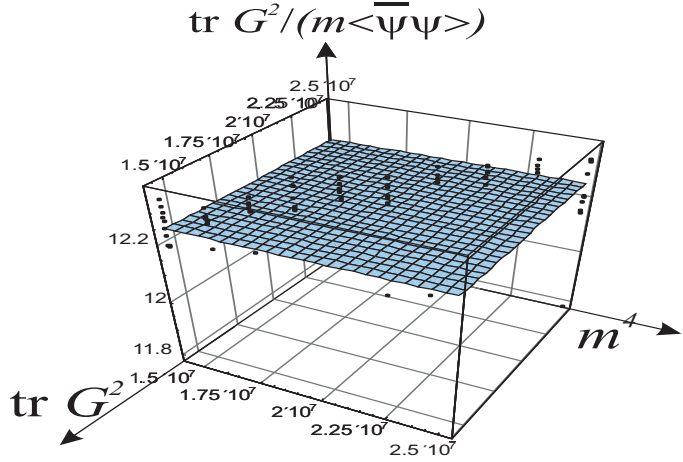}
\caption{\label{cratio}Dependence of the ratio$\frac{m
\langle\overline{q}q\rangle}{\langle\tr G^2\rangle}$ on quark mass
}
\end{center}
\end{figure}

\noindent We obtain numerically
the dimensionless ratio of the solution coefficients
$\frac{ac}{b_4}$, linearly related to the condensate ratio by
\eqref{acb}. The ratio of the condensates obtained numerically is
shown in\myfigref{cratio} for the Gubser background. Similar
pictures are obtained for the other two backgrounds. Each point in
the parameter space represents an individual ``measurement'', that
is, a solution for a $D7$-brane embedding at given gluon
condensate and quark mass, from which  the value for the quark
condensate follows. By fitting the ``experimental'' points we
estimate the value of the ratio and a statistical error margin
thereof. The mass is assumed to be the largest scale under
consideration.

We obtain numerically the following results for the dimensionless
ratio of the condensates we are looking for:
\beq-\frac{\frac{g_{YM}^2}{4\pi^2} \la \tr(G^2) \ra}{M \la
\bar{q}q \ra}=\left\{
\begin{array}{ll}
\mbox{Constable--Myers} & 12.0078\pm 0.005 \\
\mbox{Gubser} & 12.25 \pm 0.01 \\ \mbox{Liu-Tseytlin}&11.9192 \pm
0.0020
\end{array}\right.
\eeq
Comparing the results to the correct analytic value
$-\frac{\frac{g_{YM}^2}{4\pi^2} \la \tr(G^2) \ra}{M \la \bar{q}q
\ra}=12$, we see agreement with good accuracy. The obvious
universality of the three different metrics might signal that the
decoupling theorem is insensitive to the details of the IR
physics.

\section{Discussion}
Let us restate the main results of this study:
\begin{itemize}
\item We have established a universal constant value for the ratio
$\frac{m \langle\overline{q}q\rangle}{\langle\tr G^2\rangle}$ in
the holographic duality with a good precision ($0.5\%$), thus
supporting the validity of the heavy flavor decoupling in
holographic models of QCD. \item We have obtained a version of the
low-energy theorem $\int \langle T \mathcal{O}\rangle=\mathrm{dim}
(\mathcal{O})\langle\mathcal{O}\rangle$ satisfied in holography
with condensates for the pure glue sector.
\end{itemize}

In addition we also find the following results
\begin{itemize}
\item A non-trivial relation between two-point and one-point
functions $\int \langle G^2 G^2\rangle=const\langle G^2\rangle$
has been established. \item Shear and bulk viscosities have been
shown to be independent of condensates. \item The quarkonium
diffusion coefficient has been obtained both at non-zero
temperature and for a non-vanishing condensate in the Appendix A.
\end{itemize}

The significance of establishing the decoupling ratio is its
relevance to justifying phenomenological approaches to QCD via
holography.  We hope to find an analytic explanation of the
amazing agreement which appears not to be a coincidence. Here we provide
demonstration for a very important example of a statement
which relates the quark and gluon sectors. This should encourage
further development of realistic AdS/QCD constructions based on
geometries with broken scale-invariance.

\section*{Acknowledgements}
A.Z. is extremely grateful to Prof. Dr. Dietrich Habs for his
generosity, encouragement and a fruitful atmosphere of Theory vs.
Experiment interaction he has created in Garching providing a
powerful inspiration for this work. Thanks to the Organizers of
the International School on Strings and Fundamental Physics in
Munich/Garching.  A.Z. thanks Derek Teaney for correspondence on
transport coefficients. The work was also supported  by the DFG
Cluster of Excellence MAP (Munich Centre of Advanced Photonics)
(A.Z.), and by the Cluster of Excellence ``Origin and Structure of
the Universe'' (J.E.). The work of A.G. is supported in part by
the grants PICS- 07-0292165, RFBR-09-02-00308 and CRDF -
RUP2-2961-MO-09. The work of P.K. is supported in part by the
grant RFBR-09-02-00308 and by the Dynasty foundation. The work of
A.K. is supported in part by RFBR grant no. 10-02-01483 and by the
Dynasty foundation. The work of A.Z. is supported in part by the
RFBR grant 10-01-00836 supported by Ministry of Education and
Science of the Russian Federation under the contract
14.740.11.0081.

\appendix

\section{\label{quarkonia} Quarkonium
Transport in Self-Dual Background}
\subsection{Self-Dual Background at Zero Temperature}
Here we review the method of\mycite{Dusling:2008tg} for
calculating quarkonium transport properties.
The basic result of this discussion is a decoupled structure, in
which the contributions of the fermionic part of the action will
be separated from those of the gluonic part according to the
pattern \beq \mbox{meson kin.
coeff.}=\left[\mbox{\parbox{4cm}{meson mass shift\\ (D7
contribution)}}\right]\times\quad\left[\mbox{
\parbox{4cm}{two-point correlator\\
(D3 contribution)}}\right].\eeq Consider a complex field $\varphi$
of a slowly moving meson of velocity $v$, coupled to some
operators of gluonic sector, \beq\label{nonrel} L=\varphi^+ v
\partial_t \varphi+\sum_n c_n \varphi^+ \mathcal{O}_n \varphi,\eeq
where coefficients $c_n$ are defined e.g. from $D7$ action of a
dual model, which secures existence of mesons. The latter are
understood as eigenmodes of fluctuations above the classical
solution of $D7$ equations of motion. Interaction terms modify the
spectra of eigenmodes in bulk; in terms of the boundary theory
this amounts to meson mass shift. Coefficients $C_n$ are then
introduced as ``susceptibility'' of mass with regard to switching
on the operator $\mathcal{O}_n$: \beq\delta
M=-c_n\langle\mathcal{O}_n\rangle.\eeq Considering one-particle
dynamics we can obtain from\myref{nonrel} \beq
\frac{dp_i}{dt}=\mathcal{F}_i,\eeq where \beq\mathcal{F}_i=\int
d^3x \varphi^+ \nabla_i c_n\mathcal{O}_n\varphi,\eeq while
correlator of two forces is directly related to transport
coefficient \beq \kappa=\frac{1}{3}\int dt
\langle\mathcal{F}(t)\mathcal{F}(0)\rangle.\eeq One can integrate
field $\varphi$ out of these relations and obtain finally \beq
\kappa=\frac{1}{3}\int k^2d^3k c_n^2 \frac{2T}{\omega}\mathrm{Im}
\left.\langle\mathcal{O}_n\mathcal{O}_n\rangle\right|_{k},\eeq
where
\beq\left.\langle\mathcal{O}_n\mathcal{O}_n\rangle\right|_{k}=\int
d^4x \theta(t) e^{i(\omega
t-\vec{k}\vec{x})}\langle\mathcal{O}_n(x)
\mathcal{O}_n(0)\rangle.\eeq Here the contributions of flavor
dynamics and pure gluodynamics are decoupled; below we proceed in
calculating the gluodynamical part (the two-point correlator); the
coefficients $c_n$ being responsible for mass shifts are known in
literature.

\subsection{Self-Dual Background at Finite Temperature} It is
possible to obtain quarkonium diffusion and relaxation
coefficients at finite temperature and condensate, extending the
work\mycite{Dusling:2008tg}\footnote{We thank Derek Teaney for
providing us with his unpublished Notes.} to the background
of\mycite{Ghoroku:2005tf}.  This background has the metric\beq
ds^2=R^2\left(\frac{1-r^4 \pi^4
T^4}{r^2}dt^2+\frac{dx_3^2}{r^2}+\frac{dr^2}{r^2(1-r^4 \pi^4
T^4)}\right)+R^2 d\Omega^2_5,\eeq dilaton is
 \beq e^\phi=1+\frac{q}{\pi^4
T^4}\log\left(\frac{1}{1-r^4 \pi^4 T^4}\right),\eeq axion is
related to dilaton in the same way as in the zero-temperature
Liu-Tseytlin background \beq C=e^{-\phi}-1.\eeq Quarkonium
transport coefficients are quantities which feel both the
fermionic piece of the action (some embedded brane) and the
gluodynamics. From the former comes mass susceptibility to
condensate, from the latter -- correlators of interest. In
principle, it would make a good sense to work in a back-reacted
metric, however this we postpone till the method is fully
technically developed for the well-controllable
Ghoroku--Liu--Tseytlin metric.

\noindent For convenience we further use the variable \beq u=r^2
\pi^2 T^2,\eeq which lives in the interval $(0,1)$. We consider a
reduced sector of the fluctuations, namely, those of fields
$\eta^+,\eta^-,h_{11}+h_{22}$. The equations of motion are given
in Appendix A \myref{eomts}.

We see now that the problem of fields coupling to each other is
additionally burdened by presence of finite temperature. Yet
diagonalization of these equations is possible by means of the
following functional transformation \beq \left\{\begin{array}{rcl}
 \bar{\eta} ^+(u)&=&\eta ^+(u) \\
 \bar{h(u)}&=&h(u)+q \left(C_1-\pi ^2
  \log \left(1-u^2\right)\right) \eta ^+(u) \\
 \bar{\eta} ^-(u)&=&q h(u) \left(F_1-\frac{\log \left(1
 -u^2\right)}{2 \pi ^2}\right)+\eta ^-(u)
 +\\&& \hspace{.2cm}+q \left(\frac{1}{4} q
   \log ^2\left(1-u^2\right)-\frac{q C_1 \log
    \left(1-u^2\right)}{2 \pi ^2}+C_2\right) \eta ^+(u)
\end{array}\right.\eeq
Now for each of the variables we can write down an equation
similar to that for the simple dilaton modes: \beq\varphi
''(u)+\frac{u \left(u^3+6 u+4 \omega ^2+4 k^2
\left(u^2-1\right)\right)-3}{4 u^2
   \left(u^2-1\right)^2} \varphi (u)=0, \eeq
for which transport coefficient is known; we calculated it
independently, and found it to be in agreement with the previous
results\mycite{Dusling:2008tg} \beq
\frac{2\omega}{T}G_{\phi,\phi}=\pi^2 k^4 e^{-2C_\gamma k/T},\eeq
where $C_\gamma=4 \sqrt{\frac{2}{\pi }} \Gamma
\left(\frac{5}{4}\right)^2\approx 2.62$. Knowledge of
diagonalization matrix allows us to transform these results (at
$q=0$) into non-zero-condensate background: \beq
\langle\label{cmtr}\Phi_i\Phi_j\rangle=(\hat{1}+q
A)\langle\Phi_i^\prime\Phi_j^\prime\rangle_{q=0} (\hat{1}+q
A)^+,\eeq where zero-condensate solutions are rotated to
non-zero-condensate by the following rotation matrix in mode
space: \beq A=\left(\begin{array}{lll} 0&0&0\\ \pi^2& 0&0\\
0&1/2/\pi^2 &0
\end{array}\right),\eeq
and the non-perturbed matrix of finite-temperature correlators is
diagonal
 \beq \langle\Phi_i\Phi_j\rangle_{q=0}
 =\left(\begin{array}{lll}
 \langle\tr G^{+2}\tr G^{+2}\rangle&0&0\\
 0&\langle T T \rangle &0\\
0&0& \langle\tr G^{-2}\tr G^{-2}\rangle
\end{array}\right),\eeq
whence one easily gets the mesonic transport coefficient by use of
the following formula:
\beq\label{phsp}\kappa=\sum_\mathcal{O}c_\mathcal{O}^2\frac{1}{3}\frac{\pi}{2}\int
k^2 \frac{d^3k}{(2\pi)^3} \frac{2\omega}{T}
\langle\Phi_\mathcal{O}\Phi_\mathcal{O}\rangle, \eeq where the
respective mass susceptibility coefficients are obtained from
considering the fermionic fluctuations coming from the embedded
$D7$ brane piece of the action, and are defined via \beq \delta
M=-c_\mathcal{O}\langle\mathcal{O}\rangle, \eeq where $M$ refers
to the mass of quarkonium.

The correlators themselves are obtained in the following way,
which we illustrate on the example of dilaton. We consider three
domains: UV, IR and the intermediate domain (we denote the latter
QC for semiclassics, since semiclassical approximate solutions
will be valid therein). The physical limitations are infalling
boundary condition on the horizon and reflected wave in the UV,
which reduces number of unknown coefficients from 6 to 4. Then, we
have matching conditions separate for each of the modes in the
matching regions between UV and QC, an between QC and IR. This
provides additional 4 constraints, thus the system is fully
defined.
 In the UV the
general solution to EOM is \beq\phi=\frac{2 u I_2\left(2 \sqrt{u}
\sqrt{k^2-\omega ^2}\right)
   C_1}{k^2-\omega ^2}+2 u \left(k^2-\omega ^2\right) K_2\left(2
   \sqrt{u} \sqrt{k^2-\omega ^2}\right) C_2.\eeq
Taking the UV asymptotic ($u\to 0$) of $\phi$, we see that
physical boundary conditions are $C_1=B, C_2=1$, where $B$ is
related to correlator straightforwardly: \beq
\frac{2\omega}{T}G_{\phi\phi}=\frac{Im B}{\omega}.\eeq On the
contrary, expanding it for large $k$, we get the form appropriate
for matching with QC: \beq \phi=e^{-2 k \sqrt{u}} \sqrt{\pi }
k^{-29/2}   u^{-5/4}-\frac{B e^{2 k \sqrt{u}} k^{-9/2}
   u^{-5/4}}{ \sqrt{\pi }}.\eeq
The semiclassical equation has the approximate potential \beq
V_{QC}=\frac{k^2}{u\left(1-u^2\right)},\eeq which allows to obtain
the wave-functions in the standard way \beq\psi_{1,2}=\frac{e^{\pm
\int pdx}}{\sqrt{p}},\eeq where \beq p=\sqrt{V_{QC}-E}.\eeq The
semiclassical solution near $u=0$ and $u=1$ is
\beq\begin{array}{l}\displaystyle
 \phi_{QC,u\to 0}=-\frac{i e^{-2 k \sqrt{u}} \left(e^{4 k \sqrt{u}}
   A_1+A_2\right)}{\sqrt{k} \sqrt[4]{u}} \\ \\ \displaystyle   \phi_{QC,u\to 1}=
 -\frac{i e^{-\sqrt{2} k \left(\sqrt{1-u}+1\right)} \left(e^{2
   \sqrt{2} k} A_1+e^{2 k \sqrt{2(1-u)}} A_2\right)}{\sqrt{k}
   \sqrt[4]{2-2 u}}.
\end{array} \eeq
The IR solution with infalling boundary condition has only one
degree of freedom: \beq \phi_{IR}=\left(
  e^{\sqrt{2} k \sqrt{1-u} }\csc (\pi  \omega +
  e^{-\sqrt{2} k
   \sqrt{1-u}} \right)\frac{\sqrt{\pi } C}{2^{3/4} \sqrt{k}
   \sqrt[4]{1-u}}.\eeq

\noindent Equating the QC solution branches with  those of IR and
UV solutions, we get \beq \mathrm{Im} B=\pi^2 k^4 e^{-2C_\gamma
k/T},\eeq as already stated above. Taking the integral over phase
space\myref{phsp} and performing linear transformation of
correlator matrix\myref{cmtr}, we get for transport coefficient

\beq\kappa=\frac{1}{3} T^9 \frac{60\Gamma
\left(\frac{3}{4}\right)^6}{\pi ^2 \Gamma
\left(\frac{1}{4}\right)^6}\left[c_{\tr G^{+2}}(1+2q\pi^2)
+c_{T}(1+q\pi^2)+c_{\tr G^{-2}}\right],\eeq

\noindent where $c_i$ are found in\mycite{Dusling:2008tg}, $c_{\tr
G^2}=\frac{8}{5\pi}\left(\frac{2\pi}{M_0}\right)^3$,
$c_{T}=\frac{12}{5\pi}\left(\frac{2\pi}{M_0}\right)^3$, $M_0$
being the meson mass.

\section{Equations of Motion}
Here we shown the equations of motion for Liu--Tseytlin background
in the graviton, axion and dilaton sector, corresponding to the
pure glue sector on the boundary. The definitions of the fields
are contained in eqs. \myref{list1}-\myref{liutse}.

\beq\label{fsts} \left\{\begin{array}{l} \displaystyle z
\left(\left(q \omega
z^4+\omega \right)^2-32 q^2 z^6\right) \eta _+(z)+\\
\hspace{.3cm}+\left(q z^4+1\right) \left(\left(11 q z^4+3\right)
\eta_+'(z)-z \left(q z^4+1\right) \eta _+''(z)\right)=0, \\
\\\displaystyle
 32 q^2 \eta _+(z) z^6+\left(q z^4+1\right) \left(\left(q
   z^4+1\right) \left(z^2 \omega ^2+4\right) \Phi _2(z)\right.-
    \\ \hspace{.3cm}\left.-z \left(8 q \eta
   _+'(z) z^2+\left(q z^4+1\right) \left(\Phi _2'(z)+z \Phi
   _2''(z)\right)\right)\right)=0, \\ \\ \displaystyle
\left(z^2 \omega ^2+4\right) \Phi _2(z)-z
 \left(\Phi _2'(z)+z \Phi
   _2''(z)\right)=0, \\ \\ \displaystyle
\left(z^2 \omega ^2+4\right) h_{\text{xy}}(z)-z
   \left(h_{\text{xy}}'(z)+z h_{\text{xy}}''(z)\right)=0,\\
   \\ \displaystyle
-32 q^2 \eta _+(z) z^7+\left(q \omega
 z^4+\omega \right)^2 \eta
   _-(z) z- \\ \hspace{.3cm}-\left(q z^4
   +1\right) \left(8 q \Phi _2(z) z^5+\left(4 q \Phi
   _2'(z) z^5+\left(q z^4+1\right)
    \eta _-''(z)\right) z+\left(5 q
   z^4-3\right) \eta _-'(z)\right)=0.
\end{array}\right.
\eeq

\noindent Solutions for the EOM in the Liu--Tseytlin case at zero
frequency $\omega=0$ are:

\beq\label{phisol} \left(\begin{array}{l}\Phi_1\\ \Phi_2\\
\Phi_3\\ \Phi_4\\ \Phi_5\end{array}\right)=\left(
\begin{array}{l}
 C_2 \left(q z^4+1\right)^2+C_1 \left(q z^4+1\right) \\
 \frac{-q^2 C_2 z^8+C_3 z^4+C_4}{z^2} \\
 \frac{C_8 z^4+C_7}{z^2} \\
 \frac{C_{10} z^4+C_9}{z^2} \\
 q C_5-\frac{C_6 q^2+\left(q \left(q z^4+2\right) z^4+2\right) \left(4 q
   \left(C_1+C_2\right)+2 C_3\right)}{4 q \left(q z^4+1\right)}
\end{array}
\right)\eeq

\noindent Solution modes for a non-zero frequency:

 \beq\label{phisolomega}
\begin{array}{l}
\Phi_1= \frac{1}{2} q \omega ^2 K_2(z \omega ) C_1 z^6
+\frac{1}{2} \omega ^2 K_2(z \omega ) C_1 z^2, \\
\\  \Phi_2=  C_1 \left[\frac{\gamma  q  \omega ^8 z^{10}}{6144}-\frac{161 q \omega ^8
   z^{10}}{552960}+\right.\\ \left. \hspace{.3cm} + \frac{q \omega ^8 \log (z) z^{10}}{6144}+\frac{q
   \omega ^8 \log (\omega ) z^{10}}{6144}-\frac{q \omega ^8 \log (16)
   z^{10}}{92160}-\right.\\ \left.\hspace{.3cm}-\frac{q \omega ^8 \log (8) z^{10}}{27648}-\frac{q
   \omega ^8 \log (4) z^{10}}{184320}+\frac{1}{192} \gamma  q \omega ^6
   z^8-\frac{169 q \omega ^6 z^8}{23040}+\right.\\ \left. \hspace{.3cm}+\frac{1}{192} q \omega ^6 \log
   (z) z^8+\frac{1}{192} q \omega ^6 \log (\omega ) z^8-\frac{1}{960} q
   \omega ^6 \log (16) z^8 -\right.\\ \left. \hspace{.3cm}-\frac{q \omega ^6 \log (4)
   z^8}{1920}+\frac{1}{16} \gamma  q \omega ^4 z^6-\frac{17}{384} q
   \omega ^4 z^6+\frac{1}{16} q \omega ^4 \log (z) z^6
   +\right.\\ \left. \hspace{.3cm}+\frac{1}{16} q
   \omega ^4 \log (\omega ) z^6-\frac{1}{32} q \omega ^4 \log (4)
   z^6+\frac{1}{3} q \omega ^2 z^4\right] +\frac{1}{2} \omega ^2 K_2(z \omega ) C_2,\\ \\
\displaystyle \Phi_3 = \frac{1}{2} \omega ^2 K_2(z \omega ) C_7 ,\\ \\
\displaystyle \Phi_4 = \frac{1}{2} \omega ^2 K_2(z \omega ) C_9 , \\ \\
\displaystyle \Phi_5= -\frac{1}{12} q \omega ^2 C_1
z^6+\frac{1}{6} q \omega ^2 C_4 z^6-q C_1 z^4-\frac{8 q I_2(z
\omega ) C_1 z^2}{\left(q z^4+1\right) \omega
   ^2}-\frac{\omega ^2 K_2(z \omega ) C_1 z^2}{q z^4+1}+\\ \displaystyle \hspace{1cm}+\frac{4 q^2 I_2(z \omega ) C_6 z^2}{\left(q z^4+1\right) \omega ^2}+\frac{q \omega ^2
   K_2(z \omega ) C_6 z^2}{8 \left(q z^4+1\right)}.
\end{array}
\eeq

\noindent The thermal version of the Liu--Tseytlin backgrounds
leads to the following equations of motion:

\beq\label{eomts}\left\{\begin{array}{l}
 \frac{\left(u \left(u^3+6 u+4 \omega ^2+4 k^2 \left(u^2
 -1\right)\right)-3\right) \eta ^+}{4 u^2
   \left(u^2-1\right)^2}+\eta^{+''}=0, \\ \\
 -4 q \left(u^2+1\right) h(u)
  u^2+4 \left(u^2-1\right) \left(2 q u
   h'+\pi ^2 \left(u^2-1\right) \eta ^{-''}\right)
   u^2+\\ + \pi ^2 \left(u \left(u^3+6 u+4 \omega
   ^2+4 k^2 \left(u^2-1\right)\right)-3\right)
   \eta ^-=0,\\ \\
 4 \left(h'' \left(u^2-1\right)^2+2 \pi ^2 q
 \left(2 u \left(u^2-1\right) \eta
   ^{+'}-\left(u^2+1\right) \eta ^+(u)\right)\right)
   u^2+\\+ \left(u \left(u^3+6 u+4 \omega ^2+4 k^2
   \left(u^2-1\right)\right)-3\right) h=0.
\end{array}\right.\eeq


\providecommand{\noopsort}[1]{}\providecommand{\singleletter}[1]{#1}%
\providecommand{\href}[2]{#2}\begingroup\raggedright\endgroup

\end{document}